# Empirical formalism for the production of neutron-rich nuclei using fragmentation of medium heavy neutron-rich Rare Isotope Beams

**Debasis Bhowmick, Siddhartha Dechoudhury, Vaishali Naik and Alok Chakrabarti**

Variable Energy Cyclotron Centre,
1/AF, Bidhan Nagar, Kolkata 700064

**Abstract:**

We report here an empirical formalism for predicting the cross-sections of neutron-rich nuclei produced in fragmentation of relativistic $\beta$-unstable neutron-rich projectiles in the mid-mass region. The formalism is based on the abrasion-ablation picture of the projectile fragmentation reaction. It has been shown that the formalism can accurately reproduce the experimental cross-section data for nuclei produced in the fragmentation of $^{132}$Sn. The formalism, it is shown, can also reproduce the experimental cross-sections of neutron-rich nuclei produced in the fragmentation of $^{129}$Xe fairly accurately. The formalism is used to bring out the advantage in the production cross-sections of neutron-rich isotopes, especially the more neutron-rich ones; that one can expect by using unstable doubly magic neutron-rich projectile $^{132}$Sn as compared to comparatively less neutron-rich unstable projectile $^{128}$Sn and also $^{124}$Sn, the isotope which is most neutron-rich among the stable isotopes of Sn (tin). The formalism is used to predict the cross-sections of neutron-rich nuclei produced in the fragmentation reaction of unstable $^{144}$Xe and $^{146}$Cs, which are expected to be produced with sufficient intensity in the upcoming RIB facilities and in the existing RIB facilities after beam intensity upgrade.

**1. Introduction:**

To extend the limits of neutron-rich nuclei that can be produced in the terrestrial laboratories using projectile fragmentation reaction of accelerated ion beams of stable nuclei, the use of $\beta$-unstable neutron-rich projectiles (Rare Isotope Beams or RIB) has long since been considered as a viable route. However, this method has not been used widely since the intensity of these unstable beams obtainable in the presently operating RIB facilities has been typically low, hardly exceeding a few times $10^4$ particles per second (pps). The expected gain in the production cross-sections of very exotic neutron-rich nuclei because of extra neutron-richness of the projectile (because of the so called memory effect) would thus get offset by the low intensity of the neutron-rich radioactive projectile, resulting in no net gain. However, in the upcoming ISOL and PFS type of facilities, because of the improvement in the design of the accelerators delivering the primary stable beams and the beam preparation techniques, one expects to produce a number of neutron-rich RIB projectiles with three to four orders of magnitude higher intensity. This would make the production of new neutron-rich nuclei using fragmentation of unstable RIBs advantageous.

A number of empirical formalisms have been developed for estimation of production cross-sections of neutron-rich nuclei [1, 2, 3, 4, 5] produced in the fragmentation reaction. The feature that is common to all these empirical exercises is that the prediction of production cross-sections becomes less and less reliable as the nuclei become more and more neutron-rich. The present work aims at developing a simple empirical formalism that could correctly reproduce the experimental data for production of a large number of neutron-rich nuclei in the fragmentation of $^{132}$Sn, an experiment that was carried out at GSI about 10 years back [6]. The formalism is then used to predict the cross-sections of production of neutron-rich nuclei produced in fragmentation reaction of a number of projectiles in the medium heavy mass region: namely, $^{129}$Xe, $^{124,128}$Sn, $^{144}$Xe and $^{146}$Cs.

## 2. The empirical formalism:

The simple formalism presented here is developed to predict the cross-sections of only neutron-rich nuclei produced in projectile fragmentation reaction and is based on the abrasion-ablation picture. In the abrasion stage, a few nucleons are abraded. These might comprise of only protons or only neutrons or both protons and neutrons. It is assumed that because of memory effect (*N-Z* equilibration), the fragment having same *A/Z* ratio as the projectile would be produced with maximum probability at the abrasion stage. After abrasion the abraded fragment would be left in an excited state that would de-excite by evaporation of nucleons. This is known as the ablation stage. The abrasion followed by ablation would result in the final fragment or the reaction product. It is assumed that in the ablation stage only neutrons are evaporated. This is a reasonably valid assumption because of Coulomb barrier inhibiting proton evaporation and also in neutron-rich β-unstable fragments produced after abrasion the proton separation energy would likely to exceed the neutron separation energy, inhibiting proton evaporation further. Thus in our formalism the atomic number (*Z*) of the final reaction product would be decided at the abrasion stage.

We then proceed to calculate, for a given element (*Z*), the isotope that would be produced with maximum probability. For isotopes of an element of atomic number $Z_F$ produced in the fragmentation reaction, the number of protons ($N_p$) abraded is given by: $N_p = Z_p - Z_F$, where $Z_p$ is the atomic numbers of the projectile. The number of neutrons that would be abraded with maximum probability would be given by: $N_n = N_p (N_{proj} / Z_{proj})$, where $N_{proj}$ and $Z_{proj}$ are the neutron and the atomic number of the projectile respectively. Thus for a given number of protons ($N_p$) abraded, the most probable total number of nucleons abraded would be given by: $Q = N_p + N_n$. The number of neutrons lost by evaporation at the ablation stage would increase with the excitation energy of the pre-fragment left after abrasion and the excitation energy would in turn depend upon the number of nucleons abraded. It is empirically assumed that for $Q$ nucleons abraded the most probable number of neutrons lost at the ablation stage would be given by: $N_n$ (A) $= 2\sqrt{Q}$. Thus for $N_p$ protons abraded the most probable total number of nucleons that would be lost at the abrasion and ablation stages is given by $P = Q + N_n$ (A). Thus if we assume the isotopic distribution (for a given Z, the A distribution) to be Gaussian in nature, the mean of the distribution (in A/Z) would be given by:

$$\mu = (A_p - P) / Z_F \tag{1}$$

where, $A_p$ is the mass number of the projectile. The production cross-sections of different isotopes of a single element of atomic number $Z_F$ can then be expressed as:

$$\sigma (A_F) \text{ (for } Z = Z_F) = \frac{R}{\sqrt{\pi c} Z_F} exp\left(-\frac{\left(\frac{A_F}{Z_F} - \mu\right)^2}{c}\right), \tag{2}$$

where, $c$ is the width term and $R$ is a constant for a particular Z, which actually represent the area of the Gaussian distribution of fragment mass ($A_F$), for a single element of atomic number $Z_F$, that is, the total cross-section of all the fragments (isotopes) of the given element produced in the fragmentation reaction. In equation (2), one needs to know $c$ and $R$ to calculate $\sigma (A_F)$. If one makes a reasonable guess for the width $c$, $R$ can be calculated in three ways: i) use available experimental cross-section data for at least one less exotic fragment of the particular element $Z_F$ produced in the reaction, the value of $R$ so obtained can be used to calculate the cross-sections of all other fragments, especially the more exotic ones; ii) in the absence of experimental data, one can determine $R$ by estimating the cross-section of a comparatively less exotic isotope of the particular element produced in the reaction using an existing empirical formalisms [1,2] which are known to give reasonably accurate prediction for production cross-section of less exotic isotopes. One can then proceed to calculate the cross-sections of other more exotic isotopes; iii) to develop an empirical formalism to calculate the total production cross-sections of all the isotopes of a given element produced in the reaction.

An empirical formalism to calculate the value of $R$ for different proton removal channels (option iii) and the cross-section $\sigma (A_F, Z_F)$ of a particular fragment might be written in the form:

$$\sigma(A_F, Z_F) = \sigma_T \frac{a\alpha}{\sqrt{Z_F}} exp(-a(Z_P - Z_F)) \frac{1}{\sqrt{\pi c Z_F}} exp\left(-\frac{\left(\frac{A_F}{Z_F} - \mu\right)^2}{c}\right) \quad (3)$$

In equation (3): $R = \sigma_T \frac{a\alpha}{\sqrt{Z_F}} exp(-a(Z_P - Z_F))$, where

$$\alpha = exp\left(-\frac{(A_P - A_P^S)}{Z_P}\right), \qquad \text{for } A_P > A_P^S$$
$$\alpha = exp\left(-\frac{(A_P - A_P^S)}{Z_P}\right) - 1, \qquad \text{for } A_P < A_P^S, \quad (4)$$

and $\sigma_T = \sum \sigma(A, Z)$ is the total cross-section of the projectile fragmentation reaction. In our formalism we have taken the expression for $\sigma_T$ from Summerer et.al [1]:

$$\sigma_T = 450.0 \, (A_P^{1/3} + A_T^{1/3} - 2.383) \quad (5)$$

The parameter "$\alpha$" in equation (3) depends on whether the projectile is n-rich/p-rich or stable. The stability line and hence $A_P^S$ has been determined for odd Z or even Z separately using the expressions:

$$A_P^S = 1.15798 + 1.8394 \, Z_P + 0.01341 \, Z_P^2 - 6.70956\text{E-}05 \, Z_P^3 \qquad \text{for odd Z}$$
$$A_P^S = 2.32052 + 1.9139 \, Z_P + 0.01406 \, Z_P^2 - 6.81523\text{E-}05 \, Z_P^3 \qquad \text{for even Z} \quad (6)$$

## 3. Results

The calculation of production cross-sections of neutron-rich nuclei of different elements from Ru to In produced in the fragmentation of $^{132}$Sn are shown in Fig. 1A and 1B along with the experimental data [6]. Fig. 1A represents the calculations using equation (2), that is, option (i) where the value of $R$ for each element, as listed in Table 1 has been calculated using experimental cross-section of one relatively less exotic isotope of the same element. It has been found that a value of $c = 0.008$ gives best fit to the experimental data for all the six element from Ru to In. Fig. 1B shows the same results using the general formalism, that is using option (iii) as given by equations (3) to (6), using an optimised (result of fitting) value of a = 2. The calculated data for both cases are listed in Table 2 along with the experimental data which clearly shows the agreement is fairly accurate in all cases and excellent in many cases. Fig. 2A and 2B represents the same calculations for the projectile $^{129}$Xe, a neutron deficient projectile using the same values of $c = 0.008$ and a = 2. The actual cross-section values are listed in Table 3. It can be seen that the agreement between the simulation and the experimental results are quite good in the case of neutron-rich nuclei produced in the fragmentation of $^{129}$Xe [7].

The present empirical formalism can thus be considered to be reliable for prediction of cross-sections for other projectiles in this mass range. We have used the general formalism to calculate the cross-sections of neutron-rich nuclei for $^{124,128}$Sn to bring out the advantages of using more neutron-rich projectiles like $^{132}$Sn as compared to less neutron rich projectiles like $^{124,128}$Sn. The results are shown in Fig. 3. It can be clearly seen that the more exotic the product is, the more advantageous it is to use an unstable Rare Isotope Projectile Beam. The gain is two folds: in the production cross-section as well as in the signal to noise ratio because in the case of unstable projectile an exotic isotope of interest would be produced with comparatively higher cross-sections as compared to other less exotic isotopes facilitating separation and the measurement. We have also calculated the production cross-sections of neutron-rich nuclei expected to be produced in the fragmentation of $^{144}$Xe and $^{146}$Cs. Apart from $^{132}$Sn, these two neutron-rich projectiles are also expected to be produced with enough intensities in the upcoming RIB facilities or RIB facilities upgrades. The results are shown in Figs. 4 and 5.

## 4. Conclusions

The formalism developed in this work has been found to be quite suitable for predictions of cross-sections of neutron-rich exotic nuclei in the fragmentation of stable and unstable nuclei roughly in the mass range $120 < A < 150$. The formalism might not work well in other ranges, especially if the product fragment's mass is much less (say more than 20%) than the projectile's mass. This is because the formalism is based on abrasion-ablation picture that is valid ideally only for peripheral collisions in which only a few particles are abraded in the first stage of the reaction. If the collision is more central leading to loss of a large fraction of particles (nucleons) from the initial projectile, the present formalism cannot be relied upon.

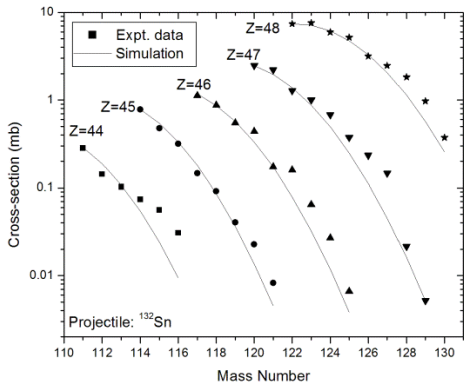 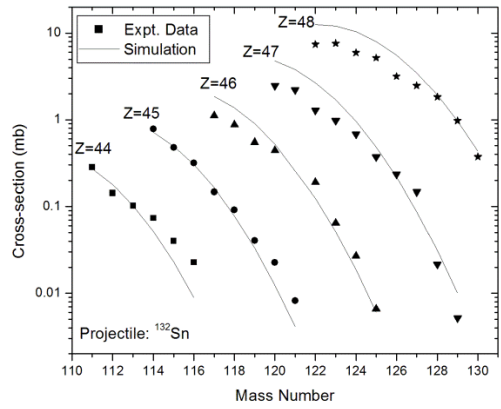

Fig. 1A  Fig. 1B

Figs.1A and 1B Comparison of simulation with experimental data of n-rich nuclei for $Z = 44 - 48$ in fragmentation of $^{132}$Sn

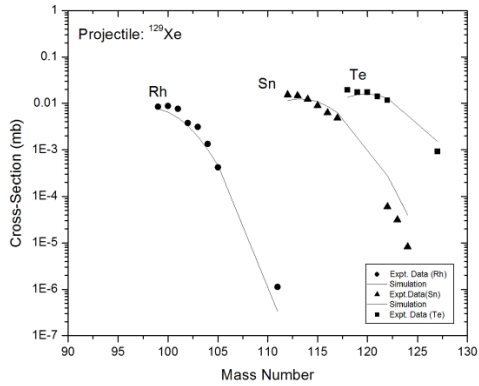 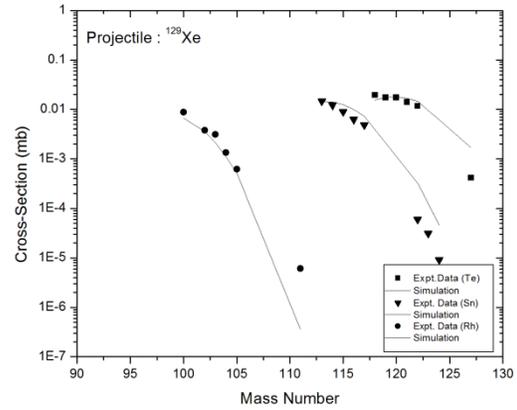

Fig. 2A  Fig. 2B

Figs. 2A and 2B. Comparison of simulation with experimental data of *n*-rich nuclei for $Z = 45, 50, 52$ in fragmentation of $^{129}$Xe

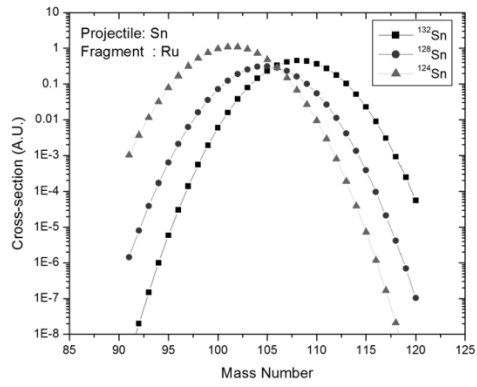

Fig. 3 Production of n-rich Ru isotopes in the fragmentation of $^{124,128,132}$Sn

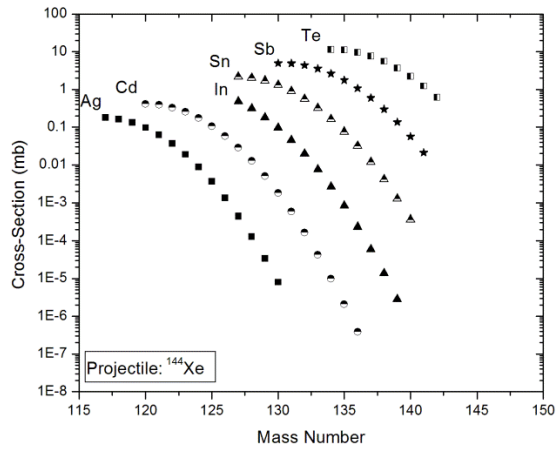

Fig. 4 Cross-sections of n-rich nuclei in fragmentation of $^{144}$Xe using present formalism

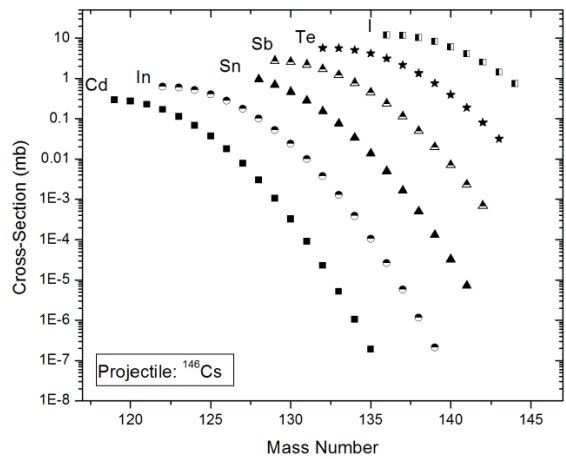

Fig. 5 Cross-sections of n-rich nuclei in fragmentation of $^{146}$Cs using present formalism

Table 1: Values of *R* as determined using *c* = .008 and experimental data of one isotope of each element from Z = 44 to 49

| Neutron-rich Projectile $^{132}$Sn | | | Neutron-deficient Projectile $^{129}$Xe | | |
|---|---|---|---|---|---|
| Product | Z | R (mb) | Product | Z | R (mb) |
| Ru | 44 | 3.27 | Rh | 45 | 50.0 |
| Rh | 45 | 8.11 | Sn | 50 | 95.0 |
| Pd | 46 | 10.86 | Te | 52 | 130.0 |
| Ag | 47 | 21.00 | | | |
| Cd | 48 | 59.30 | | | |
| In | 49 | 105.00 | | | |

Table 2: Experimental and simulated cross-sections of nuclei produced in the fragmentation of $^{132}$Sn

| Element | Z | A | $\sigma^{Expt}$ (mb) | $\sigma^{Simu}$ (mb) Using eq.1 | $\sigma^{Simu}$ (mb) Using eq. 3 |
|---|---|---|---|---|---|
| Ru | 44 | 111 | 0.2834 | 0.2833 | 0.2833 |
|   |   | 112 | 0.1428 | 0.1850 | 0.1782 |
|   |   | 113 | 0.1028 | 0.1062 | 0.1023 |
|   |   | 114 | 0.07397 | 0.05357 | 0.05158 |
|   |   | 115 | 0.05623 | 0.02374 | 0.02286 |
|   |   | 116 | 0.03077 | 0.00925 | 0.00891 |
| Rh | 45 | 114 | 0.7814 | 0.7811 | 0.7625 |
|   |   | 115 | 0.4771 | 0.5415 | 0.5125 |
|   |   | 116 | 0.3162 | 0.3318 | 0.3028 |
|   |   | 117 | 0.1468 | 0.1797 | 0.164 |
|   |   | 118 | 0.09211 | 0.08606 | 0.07853 |
|   |   | 119 | 0.04047 | 0.03641 | 0.03323 |
|   |   | 120 | 0.02276 | 0.01362 | 0.01243 |
|   |   | 121 | 0.00825 | 0.00450 | 0.00411 |
| Pd | 46 | 117 | 1.116 | 1.160 | 1.853 |
|   |   | 118 | 0.8719 | 0.8573 | 1.369 |
|   |   | 119 | 0.5471 | 0.5630 | 0.8992 |
|   |   | 120 | 0.4394 | 0.3285 | 0.5247 |
|   |   | 121 | 0.173 | 0.1703 | 0.1253 |
|   |   | 122 | 0.1594 | 0.07848 | 0.05131 |
|   |   | 123 | 0.06449 | 0.03212 | 0.01867 |
|   |   | 124 | 0.02683 | 0.01168 | 0.00603 |
| Ag | 47 | 120 | 2.471 | 2.472 | 4.771 |
|   |   | 121 | 2.214 | 1.960 | 3.783 |
|   |   | 122 | 1.28 | 1.388 | 2.68 |
|   |   | 123 | 1.000 | 0.8782 | 1.695 |
|   |   | 124 | 0.6813 | 0.4959 | 0.957 |
|   |   | 125 | 0.3728 | 0.2501 | 0.4827 |
|   |   | 126 | 0.2339 | 0.1126 | 0.2174 |
|   |   | 127 | 0.1468 | 0.04508 | 0.08742 |
|   |   | 128 | 0.0215 | 0.01619 | 0.0314 |
|   |   | 129 | 0.00518 | 0.00519 | 0.01007 |
| Cd | 48 | 122 | 7.397 | 7.390 | 12.54 |
|   |   | 123 | 7.602 | 4.342 | 12.04 |
|   |   | 124 | 5.94 | 2.289 | 10.37 |
|   |   | 125 | 5.179 | 1.082 | 8.014 |
|   |   | 126 | 3.162 | 0.4595 | 5.556 |
|   |   | 127 | 2.471 | 0.1749 | 3.456 |
|   |   | 128 | 1.828 | 1.828 | 1.929 |
|   |   | 129 | 0.973 | 0.973 | 0.9657 |
|   |   | 130 | 0.373 | 0.373 | 0.4338 |

Table 3: Experimental and simulated cross-sections of nuclei produced in the fragmentation of $^{129}$Xe

| Element | Z | A | $\sigma^{Expt}$ (mb) | $\sigma^{Simu}$ (mb) Using eq. 1 | $\sigma^{Simu}$ (mb) Using eq. 3 |
|---|---|---|---|---|---|
| Rh | 45 | 99 | 0.00842 | 0.00756 | 0.00787 |
| | | 100 | 0.00879 | 0.00646 | 0.00655 |
| | | 101 | 0.00763 | 0.00488 | 0.00522 |
| | | 102 | 0.00377 | 0.00326 | 0.00353 |
| | | 103 | 0.0031 | 0.00192 | 0.00209 |
| | | 104 | 0.00132 | 0.001 | 0.00109 |
| | | 105 | 4.15E-04 | 4.62E-04 | 5.01E-04 |
| | | 111 | 1.121E-6 | 3.316E-7 | 3.595E-7 |
| Sn | 50 | 112 | 0.01521 | 0.01153 | 0.01321 |
| | | 113 | 0.01461 | 0.01245 | 0.01438 |
| | | 114 | 0.01211 | 0.01215 | 0.01404 |
| | | 115 | 0.0089 | 0.01074 | 0.0124 |
| | | 116 | 0.0062 | 0.00859 | 0.00992 |
| | | 117 | 0.00478 | 0.00621 | 0.00717 |
| | | 122 | 6.01E-05 | 2.75E-04 | 3.17E-04 |
| | | 123 | 3.12E-05 | 1.09E-04 | 1.26E-04 |
| | | 124 | 8.20E-06 | 3.92E-05 | 4.52E-05 |
| Te | 52 | 118 | 0.01946 | 0.01357 | 0.01528 |
| | | 119 | 0.01745 | 0.01537 | 0.01731 |
| | | 120 | 0.01745 | 0.01588 | 0.01788 |
| | | 121 | 0.01405 | 0.01495 | 0.01684 |
| | | 122 | 0.0117 | 0.01284 | 0.01445 |
| | | 127 | 9.13E-04 | 0.0015 | 0.00169 |